\documentclass{ws-procs9x6}

\newcommand{\dslash}{\partial \hskip -0.5em /}
\newcommand{\vslash}{v \hskip -0.5em /}
\newcommand{\bD}{{\bf D}}

\newcommand{\xipl}{\vec{\xi\,}\hskip-0.6mm
+\hskip-0.6mm\lambda\hat{e}_3}
\newcommand{\ximl}{\vec{\xi}\hskip-0.6mm
-\hskip-0.6mm\lambda\hat{e}_3}
\newcommand{\zr}[1]{\mbox{\hspace*{#1em}}} 
\newcommand{\ID}{\mbox{{\sf 1}\zr{-0.16}\rule{0.04em}{1.55ex}\zr{0.1}}} 

\begin{document}

\title{Spin Structure Functions in Chiral Quark
Soliton Models\footnote{\uppercase{B}ased on plenary 
talk presented at the \uppercase{QCD-W}orkshop in
\uppercase{C}harlottesville, \uppercase{A}pril 2002.}}

\author{H. Weigel\footnote{
\uppercase{H}eisenberg--\uppercase{F}ellow}}

\address{Institute for Theoretical Physics\\
T\"ubingen University\\
D-72076 T\"ubingen, Germany\\
E-mail: herbert.weigel@uni-tuebingen.de}

\maketitle

\abstracts{
In this talk I review studies of hadron structure functions in
bosonized chiral quark models. Such models require regularization
and I show that the two--fold Pauli--Villars regularization scheme 
not only fully regularizes the effective action but also
leads the scaling laws for structure functions.
This scheme is consistent with other computations
of the pion structure function in that model. For the nucleon 
structure functions the present approach serves to determine 
the regularization prescription for structure functions whose 
leading moments are not given by matrix elements of local 
operators. Some numerical results are presented for the spin 
structure functions and the role of strange quarks is addressed.}

\section{Introduction}
In this talk I review the computation of hadron 
structure functions in the Nambu--Jona--Lasino (NJL) 
model\cite{Na61}. This is a particularly simple model
for quark interactions with the important feature that
the quarks can be integrated out in favor of meson
fields\cite{Eb86}. The resulting effective action for these 
mesons possesses soliton solutions\cite{Al96}. According to the 
large--$N_C$ picture\cite{tH74} of Quantum--Chromo--Dynamics (QCD) 
these solutions are interpreted as baryons. Quantization of the 
soliton then also yields baryon wave--functions in such meson 
models. The construction of hadron wave--functions is not possible 
in QCD. This represents the main obstacle for the computation of 
hadron properties from first principles. As the model adopts the 
symmetry properties of QCD, the current operators in the model 
correspond to those of QCD. As a consequence matrix elements of 
the current operators as computed in the model are physical.
In particular it is interesting to analyze the hadronic tensor 
that parameterizes the deep--inelastic--scattering (DIS) and compare
the model predictions with empirical data. This picture has led 
to interesting studies of hadron structure functions in bosonized
chiral quark models. In this talk I will present the results of
refs.\cite{We96a,We97,Schr98,We99}. These studies build up a 
consistent approach by computing the hadronic tensor (or equivalently
the forward virtual Compton amplitude) from the gauged meson action.
For the nucleon structure functions similar studies have been reported 
in refs.\cite{Di96,Wa98,Wa00}. There no attempt to compute the 
structure functions from the gauged action was made but rather 
the less convincing assumption was made that the model predictions 
for the constituent quark distribution can be identified with QCD 
quark distributions. However, I refer to those articles for a more 
expatiated presentation of numerical results.

This talk is organized as follows. In Section~2 I introduce 
the NJL model as an effective meson theory and utilize pion
properties to determine the model parameters. Section~3 describes
the subtleties for extracting the structure functions that
arise in this model from regularization. The pion structure 
function is considered as an example. In Section~4 I
review the construction of baryon states in the soliton
picture. The following Section sketches the computation of
nucleon matrix elements of the hadronic tensor and the 
extraction of the structure functions in the Bjorken limit.
Finally in Section~6 I present some numerical results
for the spin structure functions $g_1$ and $g_2$ and compare them 
to experimental data by means of the transformation to the infinite 
momentum frame and subsequent DGLAP evolution. Section~7 serves as a 
short summary.
 
\section{The NJL Model for Chiral Dynamics}

The NJL model is a quark model with a chirally invariant four 
quark interaction. Semiclassical bosonization is achieved via
introduction of effective meson fields for the
quark bilinears in that interaction. Then the quark fields can 
be integrated out by functional methods. This yields an effective 
action for meson degrees of freedom,
\begin{equation}
{\mathcal A} [S,P]=-iN_C{\rm Tr}_{\textstyle\Lambda}{\rm log}\, 
\left[i\dslash-\left(S+i\gamma_5P\right)\right]
-\frac{1}{4G}\int d^4x\, {\rm tr}\, {\mathcal V}(S,P)\, .
\label{bosact}
\end{equation}
Here ${\mathcal V}$ is a local potential respectively for the
effective scalar and pseudoscalar fields $S$ and $P$ that are 
matrices in flavor space. In the NJL model it reads
${\mathcal V}=S^2+P^2+2{\hat m}_0S$ with $\hat{m}_0$ being the
current quark mass matrix. Since the interaction is mediated by 
flavor degrees of freedom, the number of colors is merely 
a multiplicative quantity. The functional trace~(${\rm Tr}$) 
originates from integrating out the quarks and induces a non--local
interaction for $S$ and $P$. For simplicity I will only consider the 
isospin limit for up (u) and down (d) quarks: $m_{0,u}=m_{0,d}=m_0$.

A major concern in regularizing the functional (\ref{bosact}), as 
indicated by the cut--off $\Lambda$, is to maintain the chiral 
anomaly. This is achieved by splitting this functional into 
$\gamma_5$--even and odd pieces and only regulate the 
former,
\begin{eqnarray}
&&{\rm Tr}_\Lambda {\rm log}\,
\left[i\dslash-\left(S+i\gamma_5P\right)\right]
=-i\frac{N_C}{2} \sum_{i=0}^2 c_i {\rm Tr}\, {\rm log}
\left[- \bD \bD_5 +\Lambda_i^2-i\epsilon\right]
\hspace{0.4cm} \nonumber \\ &&\hspace{4.5cm}
-i\frac{N_C}{2}
{\rm Tr}\, {\rm log}
\left[-\bD \left(\bD_5\right)^{-1}-i\epsilon\right]\, ,
\label{PVreg} \\
{\rm with}\,&&\qquad
i \bD = i\dslash - \left(S+i\gamma_5P\right) 
\quad {\rm and} \quad
i \bD_5 = - i\dslash - \left(S-i\gamma_5P\right)\, .
\label{defd}
\end{eqnarray}
The double Pauli--Villars regularization renders the functional
(\ref{bosact}) finite with 
$c_0=1,\,\, \Lambda_0=0 ,\, \sum_{i=0}^2c_i=0\,.$
The $\gamma_5$--odd piece is
conditionally finite and not regularizing it properly
reproduces the chiral anomaly.  For sufficiently large $G$ one 
obtains the VEV, $\langle S\rangle\equiv m\ID$ that parameterizes 
the dynamical chiral symmetry breaking, from the gap--equation, 
\begin{equation}
\frac{1}{2G}\left(m-m_0\right)
=-4iN_C m\sum_{i=0}^2c_i
\int\frac{d^4k}{(2\pi)^4}
\left[-k^2+m^2+\Lambda_i^2-i\epsilon\right]^{-1}\, .
\label{gap}
\end{equation}
Substituting $S=\langle S\rangle=m\ID$ in 
eq.~(\ref{bosact}) shows that $m$ plays the role of a 
mass and is referred to as the constituent quark mass.

In the next step I utilize pion properties to fix the model
parameters and introduce the pion field ${\vec\pi}$ via
\begin{equation}
S+iP\gamma_5=m\, \left(U\right)^{\gamma_5} = m\, {\rm exp}
\left(i \frac{g}{m}\gamma_5\,{\vec\pi} \cdot {\vec\tau} \right)\, .
\label{SandP}
\end{equation}
Sandwiching the axial current between the 
vacuum and a single pion state yields the pion decay constant 
$f_\pi=93{\rm MeV}$ in terms of the polarization function 
$\Pi(q^2,x)$,
\begin{eqnarray}
f_\pi&=&4N_Cmg\int_0^1\, dx\, \Pi(m_\pi^2,x)
\nonumber \\ 
\Pi(q^2,x)&=&\sum_{i=0}^2 c_i\, \frac{d^4k}{(2\pi)^4i}\,
\left[k^2+x(1-x)m_\pi^2-m^2-\Lambda_i^2+i\epsilon\right]^{-2}\, ,
\label{fpi}
\end{eqnarray}
where $m_\pi=138{\rm MeV}$ is the pion mass.
The Yukawa coupling constant, $g$, is determined by the requirement
that the pion propagator has unit residuum,
\begin{equation}
\frac{1}{g^2}=4N_C \frac{d}{dm_\pi^2}\int_0^1\, dx\,
\left[m^2_\pi \Pi(m_\pi^2,x)\right]\, .
\label{yukawa}
\end{equation}
In the chiral limit ($m_\pi=0$) this simplifies to $f_\pi=m/g$. 
Finally the current quark mass is fixed from the condition that
the pole of the pion propagator is exactly at the pion mass,
\begin{equation}
m_0=4\,N_C\,m\,G\, m_\pi^2\,
\int_0^1\, dx\, \Pi(m_\pi^2,x)\, .
\label{mpi}
\end{equation}
It is also worthwhile to mention that expanding eqs.~(\ref{PVreg}) 
and (\ref{SandP}) to linear and quadratic order  
in $\vec{\pi}$ and $v_\mu$, respectively yields the proper result for
the anomalous decay $\pi^0\to\gamma\gamma$. This is the direct 
consequence of not regularizing the $\gamma_5$--odd piece.

Before discussing nucleons as solitons of the bosonized 
action~(\ref{bosact}) and the respective structure functions
it will be illuminating to first consider DIS off pions.

\section{The Compton Tensor and Pion Structure Function}

DIS off hadrons is parameterized by the hadronic tensor 
$W^{\mu\nu}(q)$ where $q$ is the momentum transmitted from
the photon to the hadron.

The tensor $W^{\mu\nu}(q)$ is obtained from the 
hadron matrix element of the current commutator by Fourier 
transformation and is parameterized in terms of 
form factors that multiply the allowed Lorentz structures. By 
pertinent projection of the hadron tensor these form factors can 
be extracted. Finally the structure functions are the leading
twist contributions of the form factors. These contributions
are obtained from computing $W^{\mu\nu}(q)$ in the Bjorken limit:
$q^2\to-\infty$ with $x=-q^2/p\cdot q$ fixed.

An essential feature of bosonized quark models is that
the derivative term in (\ref{bosact}) is formally identical to that of
a non--interacting quark model. Hence the current operator is given as
$J^\mu={\bar q}{\mathcal Q}\gamma^\mu q$, with ${\mathcal Q}$ a 
flavor matrix. Expectation values of currents are computed by
introducing pertinent sources $v_\mu$ in eq~(\ref{PVreg}) 
\begin{equation}
i\bD\, \longrightarrow \, i\bD +{\mathcal Q}\vslash
\qquad {\rm and} \qquad
i\bD_5\, \longrightarrow \, i\bD_5 -{\mathcal Q}\vslash
\label{source}
\end{equation}
and taking appropriate derivatives. In bosonized quark models it is 
convenient to start from the absorptive part of the forward 
virtual Compton amplitude\footnote{The momentum of the hadron is 
called $p$ and its spin eventually $s$.} 
\begin{eqnarray}
T^{\mu\nu}(q)&=&\int d^4\xi\, {\rm e}^{iq\cdot\xi}\,
\langle p,s|T\left(J^\mu(\xi) J^\nu(0)\right)|p,s\rangle
\nonumber \\
W^{\mu\nu}(q)&=&\frac{1}{2\pi} \mathsf{Im}\, [T^{\mu\nu}(q)]\, ,
\label{comp1}
\end{eqnarray}
because the time--ordered product is straightforwardly obtained from 
\begin{equation}
T\left(J^\mu(\xi) J^\nu(0)\right)=
\frac{\delta^2}{\delta v_\mu(\xi)\, \delta v_\nu(0)}\,
{\rm Tr}_\Lambda {\rm log}\,
\left[i\dslash-\left(S+i\gamma_5P\right)+{\mathcal Q}\,
\vslash\right]\Big|_{v_\mu=0}\, \, ,
\label{tprod}
\end{equation}
as defined from eq.~(\ref{PVreg}) with the substitution~(\ref{source}).

Pion--DIS is characterized by a single structure function,
$F(x)$. For its computation the pion matrix element in the Compton 
amplitude~(\ref{comp1}) must be specified.  For virtual pion--photon 
scattering it is obtained by expanding eqs.~(\ref{PVreg}) and 
(\ref{SandP}) to second order in both, 
${\vec\pi}$ and $v_\mu$. Due to the separation into $\bD$ and $\bD_5$ 
this calculation differs considerably from the simple evaluation of 
the `handbag' diagram. For example, isospin violating and 
dimension--five operators appear for the action~(\ref{PVreg}). 
Fortunately all isospin violating pieces cancel yielding 
\begin{eqnarray}
F(x)=\frac{5}{9} (4N_C g^2)\frac{d}{dm_\pi^2}
\left[m_\pi^2 \Pi(m_\pi^2,x)\right]\,,
\quad 0\le x\le1\,.
\label{pionf}
\end{eqnarray}
The same result is obtained by formal treatment of the divergent
handbag diagram and {\it ad hoc} regularization\cite{Fr94}.
The cancellation of the isospin violating pieces is a feature
of the Bjorken limit: insertions of the pion field on the propagator
carrying the infinitely large photon momentum can be safely ignored. 
Furthermore this propagator can be taken to be the one for 
non--interacting massless fermions. This implies that also the 
Pauli--Villars cut--offs can be omitted for this propagator. That, in 
turn, leads to the desired scaling behavior of the structure function
in this model and is a particular feature of the Pauli--Villars 
regularization. {\it A priori} it is not obvious for an arbitrary 
regularization scheme that terms of the form $q^2/\Lambda_i^2$ drop 
out in the Bjorken limit.

From eqs~(\ref{yukawa}) and~(\ref{pionf}) it is obvious that 
$F(x)=5/9$ for $0\le x\le1$
in the chiral limit ($m_\pi=0$). It must be noted that this refers 
to the structure function at the (low) energy scale of the model. To 
compare with empirical data, that are at a higher energy scale,
the DGLAG program of perturbative QCD has to be applied to $F(x)$. 
Such studies show good agreement with the data\cite{Da01}.

\section{The Nucleon as a Chiral Soliton}

Solitons are a non--perturbative 
and stationary configurations of the meson fields. To 
determine that configuration for the meson theory~(\ref{bosact})
I consider the hedgehog {\it ansatz}
$$
U_{\rm H}(\vec{r})={\rm exp}\left(i\vec{\tau}\cdot\hat{r}F(r)\right)
\quad {\rm and}\quad
\left(U_{\rm H}(\vec{r\,})\right)^{\gamma_5}=
{\rm exp}\left(i\gamma_5\vec{\tau}\cdot\hat{r}F(r)\right)
$$
for the pion field~(\ref{SandP}). The corresponding single 
particle Dirac Hamiltonian reads
\begin{equation}
h=\vec{\alpha}\cdot\vec{p} +\beta\, m\, \left[{\rm cos}F+
i\gamma_5 \vec{\tau}\cdot\hat{r}\, {\rm sin}F\right]\, .
\label{Dirac}
\end{equation}
Evaluating the action functional~(\ref{bosact}) in the 
eigenbasis of $h$ gives the energy functional\cite{Do92} in 
terms of the eigenvalues, $\epsilon_\alpha$, 
\begin{eqnarray}
E[F]&=&
\frac{N_C}{2}\left(1-{\rm sign}(\epsilon_{\rm V})\right)
\epsilon_{\rm V}
-\frac{N_C}{2}\sum_\alpha \sum_{i=0}^2 c_i 
\left\{\sqrt{\epsilon_\alpha^2+\Lambda_i^2}
-\sqrt{\epsilon_\alpha^{(0)2}+\Lambda_i^2}
\right\}
\nonumber \\ && \hspace{2cm}
+m_\pi^2f_\pi^2\int d^3r \, (1-{\rm cos}F)\, 
\label{etot}
\end{eqnarray}
for a baryon number one configuration. Here ${\rm V}$ denotes the 
unique quark level that is strongly bound by the soliton and
$\epsilon_\alpha^{(0)}$ are the eigenvalues when the soliton
is absent. The soliton profile $F(r)$ is then obtained from extremizing 
$E$ self--consistently\cite{Al96}. 

States possessing good spin and isospin quantum numbers are 
generated by taking the zero--modes to be time dependent\cite{Ad83}
\begin{equation}
U(\vec{r\,},t)=
A(t)U_{\rm H}(\vec{r\,})A^{\dag}(t)\ ,
\label{collrot}
\end{equation}
which introduces the collective coordinates $A(t)\in SU(2)$. The 
action functional is expanded\cite{Re89} up to quadratic order in 
the angular velocities 
\begin{equation} 
i\vec{\tau}\cdot\vec{\Omega}=
2A^{\dag}(t)\dot A(t)\, .
\label{angvel}
\end{equation}
The coefficient of the quadratic\footnote{A liner term does 
not arise due to isospin symmetry.} term defines the moment
of inertia\footnote{Functional integrals are evaluated using
the eigenfunctions $\phi_\alpha$ of the Dirac Hamiltonian~(\ref{Dirac})
in the background of the chiral angle $F(r)$. Thus all quantities --
like the moment of inertia -- turn into functionals of $F(r)$.}, 
$\alpha^2[F]$. Nucleon states $|N\rangle$ are obtained 
by canonical quantization of the collective coordinates, $A(t)$. This
is analogous to quantizing a rigid rotator. The eigenfunctions are the
well--known Wigner $D$--functions
\begin{equation}
\langle A|N\rangle=\frac{1}{2\pi}
D^{1/2}_{I_3,-J_3}(A)\ ,
\label{nwfct}
\end{equation}
with $I_3$ and $J_3$ being respectively the isospin and spin 
projection quantum numbers of the nucleon. This then allows to
compute matrix elements of operators in the space of the 
collective coordinates\cite{Ad83}:
\begin{equation}
\langle N |\textstyle{\frac{1}{2}}
{\rm tr}\left(\tau_a A^\dagger \tau_b A\right)|N\rangle=
-\textstyle{\frac{4}{3}}\langle N | I_a J_b|N\rangle
\quad {\rm and} \quad
\vec{\Omega}=-\vec{J\,}/\alpha^2[F]\,.
\label{nmatrix}
\end{equation}

For later use I note that the valence quark wave--function 
receives a first order perturbation
\begin{equation}
\Psi_{\rm V}(\vec{r\,},t)=
{\rm e}^{-i\epsilon_{\rm V}t}A(t)
\left\{\phi_{\rm V}(\vec{r\,})
+\frac{1}{2}\sum_{\mu\ne{\rm V}} \phi_\mu(\vec{r\,})
\frac{\langle \mu |\vec{\tau}\cdot\vec{\Omega}|{\rm V}\rangle}
{\epsilon_{\rm V}-\epsilon_\mu}\right\}\, .
\label{valrot}
\end{equation}
The moment of inertia, $\alpha^2[F]$ is 
order $N_C$, thus, upon quantization~(\ref{nmatrix}), this rotational 
correction is subleading in $1/N_C$.

Also, later in this talk I will generalize the soliton approach 
to the case of three 
flavors. The quantization of the soliton proceeds along the same 
line by taking $A(t)\in SU(3)$.  However, there is one essential 
difference: $SU(3)$ is only an approximate symmetry because the 
current mass of the strange quark differs substantially from that 
of the up and down quarks. This will be reflected
in the Hamiltonian for the collective coordinates $A(t)$ by 
an explicit flavor symmetry breaking contribution. Nevertheless
this Hamiltonian can still be diagonalized exactly by numerical
means\cite{Ya88}. As a result the eigenfunctions are not simple Wigner
$D$--functions of specific $SU(3)$ representations but rather
linear combinations thereof. These linear combinations reflect
the fact that spin--$\textstyle{\frac{1}{2}}$ baryons are no 
longer pure octet states but acquire admixture of higher 
dimensional $SU(3)$ representations. For example the nucleon 
becomes
$$
|N\rangle = |N,{\mbox{\boldmath $8$}}\rangle
+c_{1}[F]\, \frac{m_s-m}{m_s+m}\,  |N,{\mbox{\boldmath $\bar{10}$}}\rangle
+c_{2}[F]\, \frac{m_s-m}{m_s+m}\,  |N,{\mbox{\boldmath $27$}}\rangle+\ldots
\qquad\qquad
$$ 
where the constants of proportionality, $c_i$ are computed 
from the self--consistent soliton. The exact wave--functions will
be employed when studying nucleon properties in the three 
flavor model. For further details on quantizing the soliton 
in flavor $SU(3)$ I refer to the reviews\cite{We96}.

\section{Nucleon Structure Functions}

DIS off nucleons is described by four structure functions:
$F_1(x)$ and $F_2(x)$ are insensitive to the nucleon spin. 
Those associated with the components of the hadronic tensor that 
contain the nucleon spin are $g_1(x)$ and $g_2(x)$. 

As argued in section~3, the quark propagator with the infinite 
photon momentum should be taken to be the one for free and massless
fermions. Thus, it is 
sufficient to differentiate (Here $\bD$ and $\bD_5$ are those
of eq~(\ref{defd}), {\it i.e.} with $v_\mu=0$.)
\begin{eqnarray}
&&
\hspace{-0.6cm}
\frac{N_C}{4i}\sum_{i=0}^2c_i
{\rm Tr}\,\left\{\left(-\bD\bD_5+\Lambda_i^2\right)^{-1}
\left[{\mathcal Q}^2\vslash\left(\dslash\right)^{-1}\vslash\bD_5
-\bD(\vslash\left(\dslash\right)^{-1}\vslash)_5
{\mathcal Q}^2\right]\right\}
\nonumber \\ &&
+\frac{N_C}{4i}
{\rm Tr}\,\left\{\left(-\bD\bD_5\right)^{-1}
\left[{\mathcal Q}^2\vslash\left(\dslash\right)^{-1}\vslash\bD_5
+\bD(\vslash\left(\dslash\right)^{-1}\vslash)_5
{\mathcal Q}^2\right]\right\}\, ,\,\,
\label{simple}
\end{eqnarray}
with respect to the photon field $v_\mu$. 
I have introduced the $(\ldots)_5$ description
$$
\gamma_\mu\gamma_\rho\gamma_\nu
=S_{\mu\rho\nu\sigma}\gamma^\sigma
-i\epsilon_{\mu\rho\nu\sigma}\gamma^\sigma\gamma^5
\, ,\quad 
(\gamma_\mu\gamma_\rho\gamma_\nu)_5
=S_{\mu\rho\nu\sigma}\gamma^\sigma+
i\epsilon_{\mu\rho\nu\sigma}\gamma^\sigma\gamma^5
$$
to account for the unconventional appearance of axial 
sources in $\bD_5$, {\it cf.} ref.\cite{We99}. Substituting (\ref{collrot})
for the meson fields that are contained in $\bD$ and $\bD_5$,
computing the functional trace up to subleading order 
in $1/N_C$ using a basis of quark states obtained from the 
Dirac Hamiltonian~(\ref{Dirac}), yields analytical results for 
the structure functions. I refer to ref.\cite{We99} for detailed 
formulae for other structure functions and the verification of the 
sum rules that relate integrals over the structure functions
to static nucleon properties. As an example I restrain myself to
list the contribution to $g_1(x)$ which is leading order in $1/N_C$:
\begin{eqnarray}
g_1(x)&=& \frac{M_NN_C}{36i}
\Big\langle N\Big| I_3 \Big| N\Big\rangle
\int \frac{d\omega}{2\pi} \sum_\alpha \int d^3\xi
\int \frac{d\lambda}{2\pi}\, {\rm e}^{iM_Nx\lambda}
\hspace{3cm}\nonumber \\ && \hspace{-1cm}\times
\left(\sum_{i=0}^2\frac{c_i\left(\omega+\epsilon_\alpha\right)}
{\omega^2-\epsilon_\alpha^2-\Lambda_i^2+i\epsilon}\right)_{\rm P}
\Big[\phi^\dagger_\alpha(\vec{\xi\,})\tau_3
\left(1-\alpha_3\right)\gamma_5
\phi_\alpha(\xipl)
{\rm e}^{-i\omega\lambda}
\nonumber \\ && \hspace{3.9cm}
+\phi^\dagger_\alpha(\vec{\xi})\tau_3
\left(1-\alpha_3\right)\gamma_5
\phi_\alpha(\ximl)
{\rm e}^{i\omega\lambda}\Big]\, ,
\label{g1x}
\end{eqnarray}
where the subscript ($P$) indicates the pole term.

Before discussing numerical results I would like to mention the
unexpected result that the structure function entering the Gottfried sum rule 
is related to the $\gamma_5$--odd piece of the action and hence does not 
undergo regularization. This is surprising because in the parton model 
this structure function differs from the one associated with the Adler 
sum rule only by the sign of the anti--quark distribution. The latter 
structure function, however, gets regularized in the present model, 
in agreement with the quantization rules for the collective coordinates. 

\section{Numerical Results for Spin Structure Functions}

Unfortunately numerical results for the full structure functions 
in the double Pauli--Villars regularization scheme,
{\it i.e.} including the properly regularized vacuum piece are not yet 
available. However, in the Pauli--Villars regularization the axial 
charges are saturated to 95\% or more by their valence quark 
(\ref{valrot}) contributions once the self--consistent soliton 
is substituted. This provides sufficient justification to consider 
the valence quark contribution to the polarized structure functions
as a reliable approximation since {\it e.g.} the zeroth moment
of the leading structure function~$g_1$ is nothing but the axial current
matrix element. This valence quark level is that of the chiral soliton 
model and its contributions to the structure functions should not be 
confused with valence quark distributions in parton models. In general,
it should be stressed that the present model calculation yields
structure functions, {\it i.e.} quantities that parameterize the
hadronic tensor, but not (anti)--quark distributions. The latter
would require the identification of model degrees of freedom 
with those in QCD. However, here only the symmetries (namely
the chiral symmetry) and thus the current operators in the hadronic
tensor are identified.

\begin{figure}[t]
~~\parbox[l]{5.2cm}{
\psfig{figure=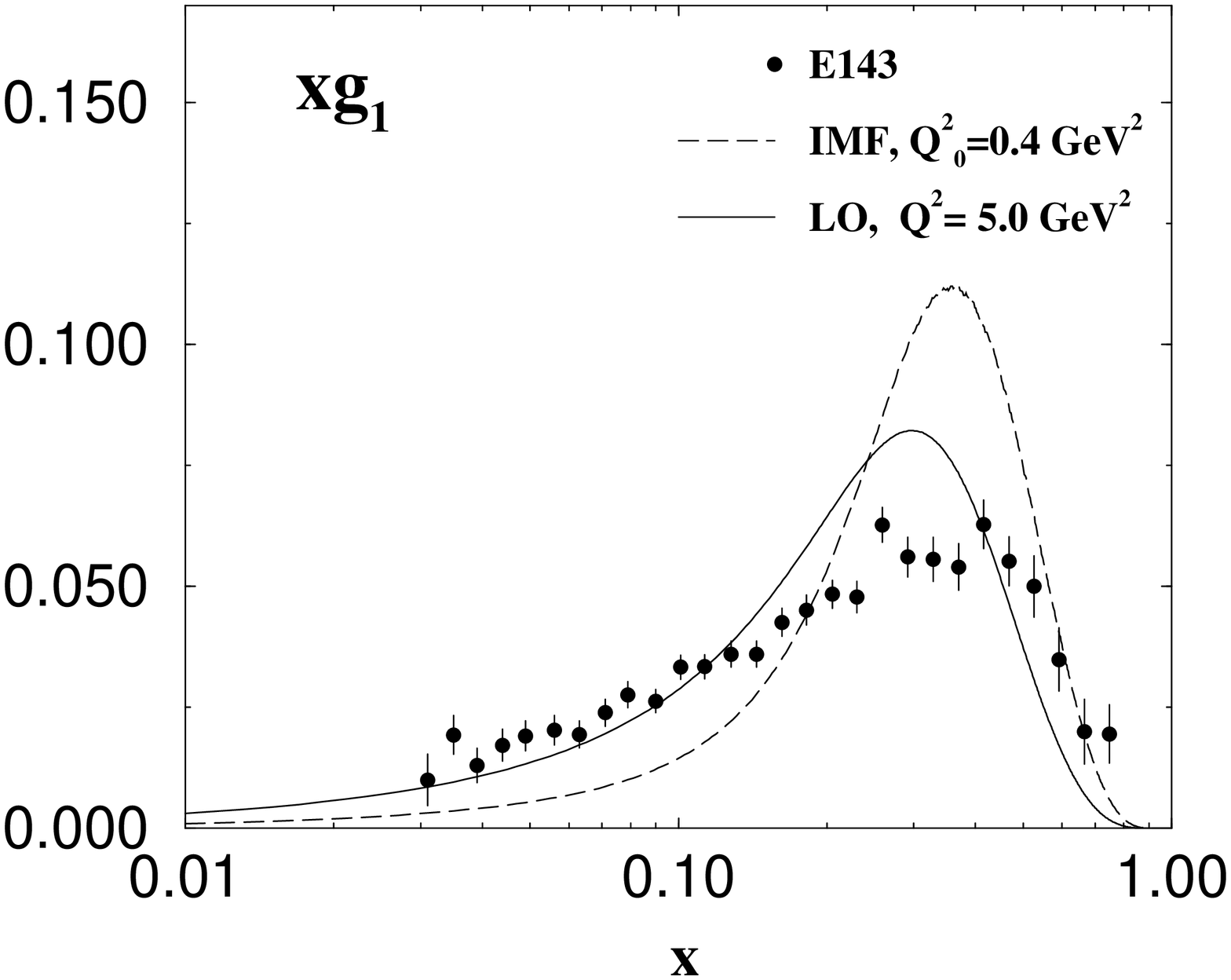,height=5.0cm,width=5.0cm,angle=270}}
~~~~~~
\parbox[r]{5.2cm}{
\psfig{figure=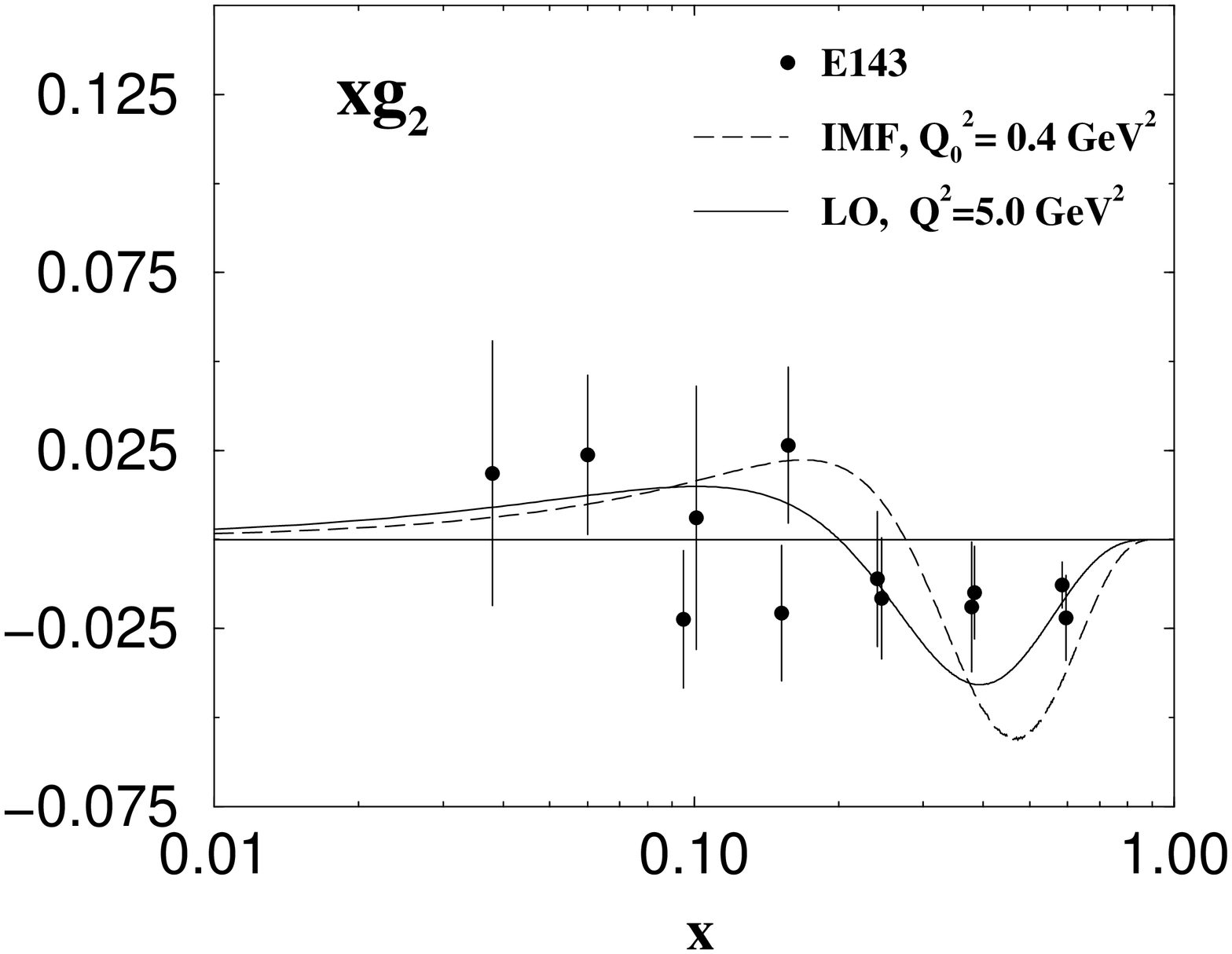,height=5.0cm,width=5.0cm,angle=270}}
\vskip-0.5cm
\caption{\label{xg12}{\sf Model predictions for the
polarized proton structure functions $xg_1$ (left panel)
and $xg_2$ (right panel). The curves labeled `RF' denote the
results as obtained from the valence quark contribution to
(\protect\ref{simple}). These undergo a projection to the infinite
momentum frame `IMF'~(\protect\ref{IMF}) and a leading order `LO'
DGLAP evolution\protect\cite{DGLAP}. 
Data are from SLAC--E143\protect\cite{Abe98}. }}
\end{figure}

\begin{figure}[t]
\psfig{figure=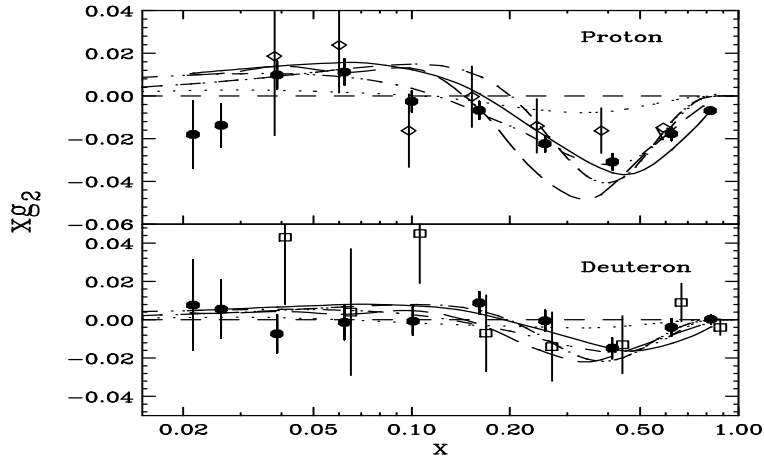,height=6.0cm,width=10.0cm}
\caption{\label{xg2w}{\sf Model predictions for the
polarized proton structure functions $xg_2$ for proton
and neutron (deuteron) and comparison with data from
E143\protect\cite{Abe98} (open diamond) and 
E155\protect\cite{An02} (open square) and their 
combination (solid circle). The full line is the 
twist--2 truncation~(\protect\ref{g2w}) of data for $g_1(x)$. 
Dashed--dotted\protect\cite{St93} and dotted\protect\cite{So96} 
lines are bag model calculations,
the short dashed lines represent the present chiral soliton 
model\protect\cite{We97} and long dashed line that of 
ref.\protect\cite{Wa00}. (This is a slightly modified figure 
from ref.\protect\cite{An02}.)}}
\end{figure}

As in the case for the pion, the model calculation yields the
nucleon structure function at a low energy scale. In addition
it must be noted that the soliton is a localized object and thus
not a representation of the Poincar\'e group. As a result the 
computation of structure functions is not frame--independent.
It is appropriate to choose the infinite momentum frame (IMF) 
not only because it makes contact with the parton model but also
because it is that frame in which the support of the structure
functions is limited to the physical regime $0\le x\le1$.
Choosing the IMF amounts to the transformation\cite{Ja81,Ga98}
\begin{equation}
f_{\rm IMF}(x)=\frac{1}{1-x}\,f_{\rm RF}(-{\rm ln}(1-x))\, ,
\label{IMF}
\end{equation}
where $f_{\rm RF}(x)$ denotes the structure function as computed
in the nucleon rest frame. So the program is two--stage, first the
transformation of the model structure function to the IMF according 
to eq~(\ref{IMF}) and subsequently the DGLAP evolution program\cite{DGLAP} 
to incorporate the ${\rm ln}|q^2|$ corrections. In the current context
it is appropriate to restrain oneself to the leading order (in $\alpha_s$)
in the evolution program because higher orders require to identify 
quark and antiquark distributions in the parton models sense. In the
present model calculation this is not possible without further 
assumptions\footnote{We assume, however, that the gluon distribution
is zero at the model scale.}. The low energy scale, $Q^2_0=0.4{\rm GeV}^2$, 
at which the model is assumed to approximate QCD has been estimated
in ref.\cite{We96a} from a best fit to the experimental data
of the unpolarized structure function, $F_2(x)$. The same boundary
value is taken to evolve the model prediction for polarized structure 
function, $g_1(x)$, in the IMF~(\ref{IMF}) to the scale $Q^2$ of 
several ${\rm GeV}^2$ at which the experimental data are available. 
For the structure function $g_2(x)$ the situation is a bit more 
complicated. First the twist--2 piece must be separated according 
to\cite{Wa77}
\begin{equation}
g_2^{WW}(x)=-g_1(x)+\int_x^1 \frac{dy}{y}\, g_1(y)
\label{g2w}
\end{equation}
and evolved analogously to $g_1(x)$ (which also is twist--2). The
remainder, $g_2(x)-g_2^{WW}(x)$ is twist--3 and is evolved according
to the large--$N_C$ scheme of ref.\cite{Ali91}. Finally, 
the two pieces are again put together at the end--point
of the evolution, $Q^2$. In figure~\ref{xg12} I compare 
the model predictions for the linearly independent polarized structure 
functions of the proton to experimental data\cite{Abe98}.
In figure~\ref{xg2w} I compare the model predictions for both  
the proton and the neutron (in form of the deuteron) not 
only to the recently accumulated data but also to other model
predictions. Surprisingly the twist--2 truncation, {\it i.e.}
eq~(\ref{g2w}) with the data for $g_1(x)$ at the right hand
side gives the most accurate description of the data. However,
also the chiral soliton model predictions reproduce the 
data well. Bag model predictions have a less pronounced 
structure.

\begin{figure}[t]
\centerline{
\psfig{figure=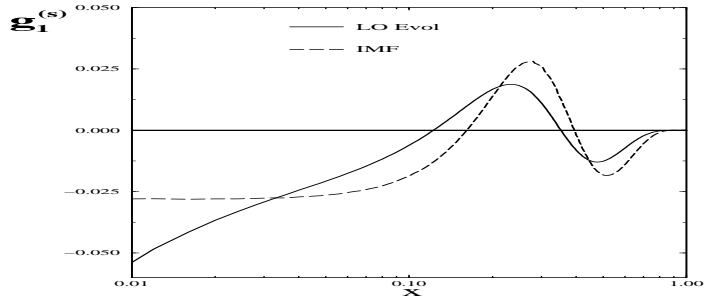,height=3.5cm,width=9.0cm}}
\caption{\label{xg1s}The strange quark contribution to the
nucleon structure function $g_1(x)$ in the infinite momentum
frame (IMF) and in leading order (LO) evolution from the
model scale $Q_0^2=0.4{\rm GeV}^2$ to $Q^2=3{\rm GeV}^2$.}
\end{figure}

As mentioned in section~4, the chiral soliton model can be
generalized to three flavors. Appropriate projection
gives a prediction for the strangeness contribution to 
structure functions\cite{Schr98}. In figure~\ref{xg1s} this 
contribution is shown for the spin structure function $g_1(x)$.
The interesting feature is that $g_1^{(s)}(x)$ has both 
positive and negative pieces. This is a nice example
showing that a small $\Delta s=\int_0^1 dx\, g_1^{(s)}(x)$  
(strange quark contribution to the nucleon spin)
does not imply that strangeness structure function itself
is small.

\section{Conclusions}

I have presented studies of the nucleon spin structure
functions in chiral soliton models. For this purpose I considered
the bosonized NJL model as a simplified model for the quark
flavor dynamics.  Although the bosonized version is an effective
meson theory, it has the interesting feature that the quark degrees 
of freedom can be traced. This is very helpful for considering 
structure functions. It turned out
that additional correlations are introduced due to the 
unavoidable regularization which is imposed in a way to respect the
chiral anomaly. Hence a consistent extraction of the nucleon structure 
functions from the Compton amplitude in the Bjorken limit leads to 
expressions that are quite different from those obtained by an 
{\it ad hoc} regularization of quark distributions in the same 
model. I also showed that within a reliable 
approximation the numerical results for the spin dependent 
structure functions agree reasonably well with the empirical 
data.

\bigskip
{\small I would like to thank the organizers for this 
worthwhile workshop. The contributions of my colleagues 
L. Gamberg, H. Reinhardt, E. Ruiz Arriola and O. Schr\"oder to 
this work are gratefully acknowledged. This work has been 
supported by the Deutsche Forschungsgemeinschaft under contract 
We 1254/3-2.}


\begin{thebibliography}{9}
\bibitem{Na61}
Y. Nambu and G. Jona--Lasinio,
\newblock Phys. Rev. {\bf 122} (1961) 345;
{\bf 124} (1961) 246.
\bibitem{Eb86}
D.~Ebert and H.~Reinhardt,
\newblock Nucl. Phys. {\bf B271} (1986) 188.
\bibitem{Al96}
R.~Alkofer, H.~Reinhardt and H.~Weigel,
Phys. Rept. {\bf 265} (1996) 139;
C. V. Christov {\it et al.}, Prog. Part. Nucl. Phys. {\bf 37} (1996) 91.
\bibitem{tH74}
G.~t`~Hooft,
\newblock Nucl. Phys. {\bf B72} (1974) 461;
{\bf B75} (1975) 461.
E.~Witten,
\newblock Nucl. Phys. {\bf B160} (1979) 57.
\bibitem{We96a}
H. Weigel, L. Gamberg and H. Reinhardt,
Mod. Phys. Lett. {\bf A11} (1996) 3021;\
Phys. Lett. {\bf B399} (1997) 287;\
L. Gamberg, H. Reinhardt and H. Weigel,
Phys. Rev. {\bf D58} (1998) 054014;\
H. Weigel, Nucl. Phys. {\bf A670} (2000) 92.
\bibitem{We97}
H. Weigel, L. Gamberg and H. Reinhardt,
Phys. Rev. {\bf D55} (1997) 6910.
\bibitem{Schr98}
O. Schr\"oder, H. Reinhardt and H. Weigel,
Phys. Lett. {\bf B439} (1998) 398.
\bibitem{We99}
H. Weigel, E. Ruiz Arriola and L. Gamberg,
Nucl. Phys. {\bf B560} (1999) 383.
\bibitem{Di96}
D. I. Diakonov {\it et al.},
Nucl. Phys. {\bf B480} (1996) 341,\
Phys. Rev. {\bf D56} (1997) 4069;\
B.~Dressler, K.~Goeke, M.~V.~Polyakov and C.~Weiss,
Eur.\ Phys. J. {\bf C14} (2000) 147.
\bibitem{Wa98}
M. Wakamatsu and T. Kubota,
Phys. Rev. {\bf D57} (1998) 5755;\
Phys. Rev. {\bf D60} (1999) 034020.
\bibitem{Wa00}
M.~Wakamatsu,
Phys.\ Lett.\ B {\bf 487} (2000) 118.
\bibitem{Fr94}
T.~Frederico and G.~A.~Miller,
Phys. Rev. {\bf D50} (1994) 210.
\bibitem{Da01}
R.~M.~Davidson and E.~Ruiz Arriola,
arXiv:hep-ph/0110291.
\bibitem{Do92}
F. D\"oring {\it et al.}, Nucl. Phys. {\bf A536} (1992) 548.
\bibitem{Ad83}
G.~S. Adkins, C.~R. Nappi, and E.~Witten,
\newblock Nucl. Phys. {\bf B228} (1983) 552.
\bibitem{Re89}
H.~Reinhardt,
Nucl. Phys. {\bf A503} (1989) 825.
\bibitem{Ya88}
H. Yabu and K. Ando,
Nucl. Phys. {\bf B301} (1988) 601.
\bibitem{We96}
H.~Weigel, Int. J. Mod. Phys. {\bf A11} (1996) 2419;\
J. Schechter and H. Weigel, arXiv:hep-ph/9907554.
\bibitem{Ja81}
R.~L.~Jaffe,
Annals Phys.\  {\bf 132} (1981) 32.
\bibitem{Ga98}
L. Gamberg, H. Reinhardt and H. Weigel,
Int. J. Mod. Phys. {\bf A13} (1998) 5519.
\bibitem{DGLAP}
G. Altarelli, P. Nason and G. Ridolfi, Phys. Lett. {\bf B320} (1994) 152;
{\bf B325} (1994) 538 (E).
\bibitem{Wa77}
S.~Wandzura and F.~Wilczek,
Phys.\ Lett.\ B {\bf 72} (1977) 195.
\bibitem{Ali91}
A. Ali, V. M. Braun and G. Hiller, Phys. Lett. {\bf B266} (1991) 117.
\bibitem{Abe98}
K.\ Abe {\it et al.},
Phys. Rev. {\bf D58} (1998) 112003.
\bibitem{An02}
P.~L.~Anthony {\it et al.}, 
arXiv:hep-ex/0204028.
\bibitem{St93}
M.~Stratmann,
Z.\ Phys.\ C {\bf 60} (1993) 763.
\bibitem{So96}
X.~Song,
Phys.\ Rev.\ D {\bf 54} (1996) 1955.
\end{thebibliography}
\end{document}